\documentclass[conference]{IEEEtran}
\usepackage[latin1,utf8]{inputenc}

\usepackage[
  pass,% keep layout unchanged 
]{geometry}

\usepackage{amsmath,bbm,comment,subfigure}

\usepackage{graphicx}
\usepackage{acronym}
\usepackage{changes}
\usepackage{balance}
\usepackage{lipsum}% <- For dummy text
\definechangesauthor[name={Authors}, color=blue]{authors}
\setremarkmarkup{(#2)}

\usepackage{acronym}
\usepackage[acronym]{glossaries}
\usepackage{glossary-mcols}
\usepackage{url}
\usepackage{booktabs}
\def\tallafigura{0.54} %0.54
\def\tallafigurabig{0.56} %0.57
\usepackage{flushend}

\usepackage{amsmath}

\usepackage{array}
\newcolumntype{L}[1]{>{\raggedright\let\newline\\\arraybackslash\hspace{0pt}}m{#1}}
\newcolumntype{C}[1]{>{\centering\let\newline\\\arraybackslash\hspace{0pt}}m{#1}}
\newcolumntype{R}[1]{>{\raggedleft\let\newline\\\arraybackslash\hspace{0pt}}m{#1}}

\makeatletter
\def\BState{\State\hskip-\ALG@thistlm}
\makeatother

\newcommand{\network}{\textit{Network} }
\newcommand{\deployment}{\textit{Deployment} }
\newcommand{\service}{\textit{Service} }
\newcommand{\bundle}{\textit{Bundle} }
\newcommand{\resources}{\textit{Resources} }
\newcommand{\assignments}{\textit{Assignments} }
\newacronym{AAA}{AAA}{Authentication, Authorization, and Accounting}
\newacronym{RSC}{RSC}{Recursive  Systematic Convolutional}
\newacronym{LLR}{LLR}{Log-Likelihood Ratio}
\newacronym{FS}{FS}{Functional Split}
\newacronym{BBU}{BBU}{Base Band Unit}
\newacronym{COTS}{COTS}{Commercial off-the-shelf }
\newacronym{VNF}{VNF}{Virtualized Network Function}
\newacronym{VNF FG}{VNF FG}{VNF Forwarding Graph}
\newacronym{NFV}{NFV}{Network Function Virtualization}
\newacronym{GPP}{GPP}{General Purpose Processor}
\newacronym{vEPC}{vEPC}{virtual Evolved Packet Core}
\newacronym{LTE}{LTE}{Long Term Evolution}
\newacronym{uRLLC}{uRLLC}{Ultra-Reliable Low-Latency Communications}
\newacronym{eMBB}{eMBB}{enhanced Mobile BroadBand}
\newacronym{mMTC}{mMTC}{massive Machine Type Communications}

\newacronym{C-EPC}{C-EPC}{Cloud-EPC}
\newacronym{EPCaaS}{EPCaaS}{EPC as a Service}
\newacronym{TDD}{TDD}{Time Division Duplex}
\newacronym{UE}{UE}{User Equipment}
\newacronym{HARQ}{HARQ}{Hybrid Automatic Repeat-Request}
\newacronym{PRB}{PRB}{Physical Resource Blocks}
\newacronym{MCS}{MCS}{Modulation and Coding Scheme}
\newacronym{CQI}{CQI}{Channel Quality Indicator}
\newacronym{DC}{DC}{Dedicated Core}
\newacronym{RR}{RR}{Round Robin}
\newacronym{G}{G}{Greedy}
\newacronym{VPN}{VPN}{Virtual Private Network}
\newacronym{MPLS}{MPLS}{Multiprotocol Label Switching}
\newacronym{OWL}{OWL}{Web Ontology Language}

\newacronym{FDD}{FDD}{Frequency Division Duplex}
\newacronym{OFDM}{OFDM}{Orthogonal Frequency Division Multiplexing}

\newacronym{VM}{VM}{Virtual Machine}
\newacronym{PDCP}{PDCP}{Packet Data Convergence Protocol}
\newacronym{MAC}{MAC}{Medium Access Control}
\newacronym{RLC}{RLC}{Radio Link Control}
\newacronym{RRC}{RRC}{Radio Resource Control}
\newacronym{AM}{AM}{Acknowledged Mode}
\newacronym{UM}{UM}{Unacknowledged Mode}
\newacronym{TM}{TM}{Transparent Mode}
\newacronym{MIMO}{MIMO}{Multiple Input Multiple Output}
\newacronym{MISO}{MISO}{Multiple Input Single Output}
\newacronym{SIMO}{SIMO}{Single Input Multiple Output}
\newacronym{SISO}{SISO}{Single Input Single Output}
\newacronym{MCC}{MCC}{Mobile Country Code}
\newacronym{MNC}{MNC}{Mobile Network Code}
\newacronym{S-TMSI}{S-TMSI}{Shortened Temporary Mobile Subscriber Identity}
\newacronym{IMSI}{IMSI}{International Mobile Subscriber Identity}
\newacronym{DRB}{DRB}{Dedicated Radio Bearer}
\newacronym{GUMMEI}{GUMMEI}{Globally Unique MME Identity}

\newacronym{PCI}{PCI}{Physical-layer Cell Identity}
\newacronym{ROHC}{ROHC}{Robust Header Compression}
\newacronym{SN}{SN}{Sequence Number}
\newacronym{RAR}{RAR}{Random Access Response}
\newacronym{C-RNTI}{C-RNTI}{Cell Radio Network Temporary Identifier}
\newacronym{BSR}{BSR}{Buffer Status Report}
\newacronym{DRX}{DRX}{Discontinuous Reception}
\newacronym{PHR}{PHR}{Power Head Room}
\newacronym{PUSCH}{PUSCH}{Physical Uplink Shared Channel}
\newacronym{ADM}{ADM}{Activation/Deactivation MAC}
\newacronym{GP}{GP}{Gap Period}

\newacronym{RE}{RE}{Resource Element}
\newacronym{RB}{RB}{Resource Block}
\newacronym{REG}{REG}{Resource Element Group}
\newacronym{CSRS}{CSRS}{Cell-Specific Reference Signal}
\newacronym{IFFT}{IFFT}{Inverse Fast Fourier Transform}
\newacronym{OFDMA}{OFDMA}{Orthogonal Frequency Division Multimple Access}
\newacronym{CRC}{CRC}{Cyclic Redundancy Check}
\newacronym{SFC}{SFC}{Service Function Chain}

\newacronym{eNB}{eNB}{Evolved NodeB}
\newacronym{RAN}{RAN}{Radio Access Network}
\newacronym{ARQ}{ARQ}{Automatic Repeat reQuest}
\newacronym{NAS}{NAS}{Non-Access Stratum}
\newacronym{MME}{MME}{Mobility Management Entity}
\newacronym{MIB}{MIB}{Master Information Block}
\newacronym{SIB}{SIB}{System Information Block}
\newacronym{RSRP}{RSRP}{Reference Signal Received Power}
\newacronym{RAT}{RAT}{Radio Access Technologie}
\newacronym{ACK}{ACK}{Acknowledge}
\newacronym{NACK}{NACK}{Negative acknowledge}
\newacronym{PDCCH}{PDCCH}{Physical Downlink Control Channel}
\newacronym{SAW}{SAW}{Stop and Wait}
\newacronym{TTI}{TTI}{Transmission Time Interval}
\newacronym{RRH}{RRH}{Radio Remote Head}
\newacronym{SNIR}{SNIR}{Signal-to-Noise-plus-Interference Ratio}
\newacronym{WCET}{WCET}{Worst Case Execution Time}
\newacronym{GPC}{GPC}{General Purpose Computer}
\newacronym{KPI}{KPI}{Key Performance Indicator}
\newacronym{OAI}{OAI}{Open Air Interface}
\newacronym{IMS}{IMS}{IP Multimedia Subsystem}
\newacronym{vIMS}{vIMS}{virtual IP Multimedia Subsystem}
\newacronym{EPC}{EPC}{Evolved Packet Core}
\newacronym{SDN}{SDN}{Software Defined Network}
\newacronym{C-RAN}{C-RAN}{Centralized-RAN}
\newacronym{OS}{OS}{Operating System}
\newacronym{TB}{TB}{Transport Block}
\newacronym{TBS}{TBS}{Transport Block Size}
\newacronym{QCI}{QCI}{QoS Channel Indicator}

\newacronym{GPU}{GPU}{Graphics Processing Unit}
\newacronym{CPU}{CPU}{Central Processing Unit}

\newacronym{SDU}{SDU}{Service Data Unit}
\newacronym{CBS}{CBS}{Code Block Size}
\newacronym{CB}{CB}{Code Block}
\newacronym{SPMD}{SPMD}{Single Program Multiple Data}
\newacronym{SIMD}{SIMD}{Single Instruction Multiple Data} 

\newacronym{SINR}{SINR}{Signal-to Interference Noise Ratio}
\newacronym{CO}{CO}{Central Office}
\newacronym{CA}{CA}{Carrier Aggregation}
\newacronym{SRS}{SRS}{Sound Reference Signal}
\newacronym{SC-OFDMA}{SC-OFDMA}{Single Carrier - Orthogonal Frequency Division Multiple Access}
\newacronym{FPGA}{FPGA}{Field-Programmable Gate Array}
\newacronym{TA}{TA}{Time Advancing}
\newacronym{CoMP}{CoMP}{Coordinated Multi-point}
\newacronym{NPRB}{NPRB}{Number of Physical Resource Blocks}
\newacronym{RTT}{RTT}{Round Trip Time}
\newacronym{CPRI}{CPRI}{Common Public Radio Interface}
\newacronym{CBR}{CBR}{Constant Bit Rate}
\newacronym{NRB}{NRB}{Number of Resource Blocks}
\newacronym{BJF}{BJF}{Biggest Job First}
\newacronym{EDF}{EDF}{Earliest Deadline First}
\newacronym{FCFS}{FCFS}{First-come, First-served}
\newacronym{PSTN}{PSTN}{Public Switched Telephone Network}
\newacronym{ETSI}{ETSI}{European Telecommunications Standards Institute}
\newacronym{vBBU}{vBBU}{virtualized BBU}
\newacronym{vRAN}{vRAN}{virtualized RAN}
\newacronym{IoT}{IoT}{Internet of Things}
\newacronym{B2B}{B2B}{Business to Business}
\newacronym{B2C}{B2C}{Business to Customer}
\newacronym{QoE}{QoE}{Quality of Experience}
\newacronym{QoS}{QoS}{Quality of Service}
\newacronym{VNO}{VNO}{Virtual mobile Network Operator}
\newacronym{SLA}{SLA}{Service Level Agreement}
\newacronym{VRRM}{VRRM}{Virtual Radio Resource Management}
\newacronym{KVM}{KVM}{Kernel-based Virtual Machine}
\newacronym{LXC}{LXC}{Linux Containers}
\newacronym{PS}{PS}{Processor Sharing}
\newacronym{eCPRI}{eCPRI}{evolved CPRI}
\newacronym{RoE}{RoE}{Radio over Ethernet}
\newacronym{PAPR}{PAPR}{Peak-to-average power ratio}
\newacronym{SC-FDMA}{SC-FDMA}{Single Carrier Frequency Division Multiple Access}
\newacronym{AGC}{AGC}{Automatic Gain Control}
\newacronym{PMD}{PMD}{Polarization Mode Dispersion}
\newacronym{ADC}{ADC}{Analogic-Digital Converter}

\newacronym{IQ}{IQ}{In-Phase Quadrature}
\newacronym{xRAN}{xRAN}{extensible Radio Access Network}
\newacronym{ISI}{ISI}{Inter-symbol interference}
\newacronym{FFT}{FFT}{Fast Fourier Transform}
\newacronym{IPC}{IPC}{Inter process communication}
\newacronym{CCDU}{CCDU}{Channel Coding Data Unit}
\newacronym{CC}{CC}{Channel Coding}
\newacronym{gNB}{gNB}{next-Generation Node B}
\newacronym{EUTRAN}{EUTRAN}{Evolved Universal Terrestrial Radio Access Network}
\newacronym{SCTP}{SCTP}{Stream Control Transmission Protocol}
\newacronym{NR}{NR}{New Radio}
\newacronym{NF}{NF}{Network Function}
\newacronym{CU}{CU}{Central Unit}
\newacronym{DU}{DU}{Distributed Unit}
\newacronym{NGC}{NGC}{Next Generation Core}
\newacronym{DL}{DL}{down-link}
\newacronym{UL}{UL}{up-link}
\newacronym{LJF}{LJF}{Largest Job First}
\newacronym{RANaaS}{RANaaS}{RAN as a Service}
\newacronym{NaaS}{NaaS}{Network as a Service}
\newacronym{NS}{NS}{Network Service}
\newacronym{FG}{FG}{Forwarding Graph}
\newacronym{VNFC}{VNFC}{VNF Component}
\newacronym{MANO}{MANO}{Management and Orchestration}
\newacronym{FIFO}{FIFO}{First In Firs Out}
\newacronym{NFVI}{NFVI}{NFV Infrastructure}
\newacronym{NFVO}{NFVO}{NFV Orchestrator}
\newacronym{PoP}{PoP}{Point of Presence}
\newacronym{NAT}{NAT}{Network Address Translation}
\newacronym{CDN}{CDN}{Content Delivery Network}
\newacronym{VNFM}{VNFM}{VNF Manager}
\newacronym{EM}{EM}{Element Management}
\newacronym{VIM}{VIM}{Virtualised Infrastructure Manager}

\newacronym{e2e}{e2e}{end-to-end}

\newacronym{AMF}{AMF}{Access and Mobility Management Function}
\newacronym{SMF}{SMF}{Session Management Function}
\newacronym{UPF}{UPF}{User Plane Function}
\newacronym{PCF}{PCF}{Policy Control Function}
\newacronym{UDM}{UDM}{Unified Data Management}
\newacronym{NRF}{NRF}{NF Repository Function}
\newacronym{AUSF}{AUSF}{Authentication Server Function}
\newacronym{API}{API}{Application Programming Interface}
\newacronym{HSS}{HSS}{Home Subscriber Server}
\newacronym{PCRF}{PCRF}{Policy and Charging Rules Function}
\newacronym{SOA}{SOA}{Software-Oriented Architecture}
\newacronym{AKA}{AKA}{Authentication and Key Agreement}
\newacronym{AF}{AF}{Application Function}
\newacronym{NEF}{NEF}{Network Exposure Function}
\newacronym{NSSF}{NSSF}{Network Slice Selection Function}
\newacronym{NSSP}{NSSP}{Network Slice Service Profile}
\newacronym{NSSAI}{NSSAI}{Network Slice Selection Assistance Information}
\newacronym{NSST}{NSST}{Network Slice Subnet Template}

\newacronym{NSSI}{NSSI}{Network Slice Subnet Instance}

\newacronym{NSS}{NSS}{Network Slice Subnet}
\newacronym{NSC}{NSC}{Network Slice Customer}
\newacronym{NSP}{NSP}{Network Slice Provider}
\newacronym{CSC}{CSC}{Communication Service Customer}
\newacronym{CSP}{CSP}{Communication Service Provider}
\newacronym{SST}{SST}{Slice/Service Type}
\newacronym{SD}{SD}{Slice Diferenciator}
\newacronym{USRP}{USRP}{UE Router Selection Policy}
\newacronym{S-NSSAI}{S-NSSAI}{Single Network Slice Selection Assistance Information}
\newacronym{ONISTT}{ONISTT}{Open Net-centric Interoperability Standards for Training and Testing}
\newacronym{KB}{KB}{Knowledge Base}
\newacronym{NSI}{NSI}{Network Slice Instance}

\newacronym{VF}{VF}{Virtual Function}
\newacronym{VFC}{VFC}{Virtual Function Component}
\newacronym{CR}{CR}{Complex Resource}
\newacronym{PNF}{PNF}{Physical Network Function}
\newacronym{CP}{CP}{Connection Point}
\newacronym{VL}{VL}{Virtual Link}
\newacronym{SDC}{SDC}{Service Design and Creation}
\newacronym{ONAP}{ONAP}{Open Network Automation Platform}
\newacronym{VID}{VID}{Virtual Infrastructure Deployment}
\newacronym{VSP}{VSP}{Vendor Software Product}

\newacronym{WEF}{WEF}{Wireless Edge Factory}

\newacronym{DP}{DP}{Data Plane}
\newacronym{ECOMP}{ECOMP}{Enhanced Control Orchestration Management and Policy}

\newacronym{AAI}{AAI}{Active and Available Inventory}
\newacronym{SDNC}{SDNC}{Software Defined Network Controller}
\newacronym{SO}{SO}{Service Orchestrator}
\newacronym{APPC}{APPC}{Application Controller}
\newacronym{DCAE}{DCAE}{Data Collection Analytics and Events}
\newacronym{OOF}{OOF}{ONAP Optimization Framework}
\newacronym{OSS}{OSS}{Operation Support System}
\newacronym{BSS}{BSS}{Business Support System}
\newacronym{SOCKS}{SOCKS}{Secured Over Credential-based Keberos}
\newacronym{VVP}{VVP}{VNF Validation Program}

\newacronym{VLM}{VLM}{Vendor License Model}

\newacronym{CUPS}{CUPS}{Control User Plane Separation}

\newglossarystyle{modsuper}{%
  \glossarystyle{super}%

}

\makeglossaries

\begin{document}

\title{Automating the deployment of 5G Network Slices with ONAP}

\author{\IEEEauthorblockN{Veronica Quintuna Rodriguez\IEEEauthorrefmark{1}, Fabrice Guillemin\IEEEauthorrefmark{1} and Amina Boubendir\IEEEauthorrefmark{2}}\\ 
\IEEEauthorblockA{\IEEEauthorrefmark{1}Orange Labs, 2 Avenue Pierre Marzin,  22300 Lannion, France} \IEEEauthorrefmark{2}Orange Labs, 44 Avenue de la République, 92320 Ch\^atillon, France\\ \{veronica.quintunarodriguez, fabrice.guillemin, amina.boubendir\}@orange.com}

%\author{\IEEEauthorblockN{Veronica Quintuna Rodriguez, Fabrice Guillemin and Amina Boubendir} \\
%\IEEEauthorblockA{Orange Labs,France} 
%\{veronica.quintunarodriguez, fabrice.guillemin, amina.boubendir\}@orange.com}

\maketitle  
\begin{abstract}
 Open Network Automation Platform (ONAP) is a carrier grade  platform for automatically deploying and managing Virtualized Network Functions. In this paper, we address the deployment of network slices in order to come up with a model that is compatible with ONAP. We analyze various types of network slice ontology presented in the framework of 5G standardization bodies and we propose an ONAP-compatible model on the basis of which we illustrate the design, onboarding, instantiation and distribution of a network slice. We concretely define and deploy a  network slice implementing a private and customized mobile core network. The achieved results not only make true NFV and 5G promises, notably those referring to on-demand networks, service customization and time-to-market acceleration, but they open the door to the deployment of private tailored cloud-native 5G networks.  
\end{abstract}


\section{Introduction}

The concept of network slicing has become central in the design and deployment of 5G networks since it enables network operators to meet the specific service requirements expressed by vertical markets, such as real-time objectives, reliability and guaranteed \glspl{SLA}. 3GPP has notably introduced three macroscopic quality-oriented  network slices namely \gls{eMBB}, \gls{mMTC}, and \gls{uRLLC}. Beyond these three macro-service classes, network slicing can be seen as a means of dedicating network resources to specific purposes and/or customers. This approach takes benefit from new possibilities offered by virtualization technologies.

Network slicing based on virtualization can in fact be dated back to GENI initiative ~\cite{GENIBook2016} and notably the concepts presented by Feamster \emph{et al}~\cite{feamster2007lease}. While network slicing was initially introduced to overcome the ossification of the Internet (e.g., via the implementation of new network protocols inside network slices), the emergence of \gls{NFV} \cite{ETSI_site} and \gls{SDN} as well as the massive deployment of Cloud infrastructures have radically changed the scope of network slicing. 

As a matter of fact, one of the unprecedented innovations brought by virtualization technologies in the design of networks is the separation between network functions and their hosting hardware. Virtual Network Functions \glspl{VNF} can thus be instantiated, on the fly, on common hardware and more generally on Cloud platforms. A network slice can then be defined as an independent virtual network running on top of a shared and/or dedicated infrastructure fulfilling specific performance requirements. A network slice is in fact a polymorphic entity which can be as simple as a \gls{VPN} with a given \gls{SLA} in terms of bandwidth and latency or a much more complex object embedding \glspl{VNF} with more or less stringent isolation and performance requirements.

Network operators can actually take advantage of \gls{NFV} and network slicing for offering portions of their networks to meet the needs of vertical industries as health-care,  automotive, financial, etc. Network slicing is thus the capability of building packages of \glspl{VNF} or more specifically Service Functions Chains \glspl{SFC} dedicated to  specific purposes. For instance, a virtualized end-to-end mobile network including both core and radio access networks (with private antennas) can be composed and dedicated to a given company for setting up a private mobile network. Such a network slice can appear as an enriched VPN with a radio component. The major innovation of network slicing is then the flexibility, adaptability and time-to-market acceleration when deploying customized networks. 

While both \gls{NFV} and \gls{SDN} respectively enable running network functions as applications and decoupling control and data functions, none of them manages the entire life-cycle of VNFs. To address this need, \gls{ONAP} initiative was launched in 2017 promising automation of network services deployment as chains of VNFs. ONAP offers a number of features involved in the service design and life-cycle management, all working together~\cite{ONAPLF}. For network operators, \gls{ONAP} concretely enables the orchestration, control and operation of end-to-end network services as network slices. ONAP also promises advanced monitoring and analytic techniques to improve legacy network services. 

In this paper, we address the problem of automating the deployment of network slices via \gls{ONAP}. We present a unified network slicing model that gathers both ONAP software entities and 3GPP network architectures. On the basis of the proposed model, we present how to design and run a network slice that implements a private mobile core network to be dedicated to a given company. In addition, we identify some limitations in the current ONAP distribution and propose solutions to overcome these shortcomings. 

We concretely present the design, on-boarding, instantiation and distribution of a fully virtualized 5G network slice which involves private core network functions and a shared access network. The exposed network slice is based on a complete separation of the data and control planes as recommended by 3GPP and known as \gls{CUPS}.
The deployed use-case makes true the 5G promises notably those referring to on-demand services and reduction of commercialization cycles. Results show that network slices can be instantiated on the fly and answer emerging markets needs.

This paper is organized as follows: In Section~\ref{related}, we review some definitions of network slicing given by the Industry and standardization bodies. In Section~\ref{ontology}, we present the proposed ontology approach and show that ontology provides models that are very relevant in the context of network slicing. In Section~\ref{onap_ontology}, we present a unified model compatible with ONAP. This model is illustrated in Section~\ref{example} in the case of a mobile core network deployed on demand. Some concluding remarks are presented in Section~\ref{conclusion}.

%A set VNFs together associated with storage resources and a virtual private network (VPN) offering transport facilities between all components is a network slice \textbf{ref}. Such a network slice can be viewed as an enriched VPN, with a radio component. This is definitely a major innovation with respect to the current state of of the art in terms of VPNs.
  %This leads us to develop an ontology of slice.

 %Various network entities can be virtualized, from elementary functions such as  firewalls up to very complex \gls{VNF} chains as 5G networks. 
 
% \textbf{ref. 3GPP}, some with stringent requirements in terms of  latency and bandwidth, such as Radio Access Network (RAN) functions. For instance, OAI RAN \textbf{OAI} is a complete software suite capable of running in virtual machines or containers all RAN functions and connecting commercial smartphones to the OAI core; this builds a complete virtualized technical mobile chain.

%http://selfnet.inf.um.es/nsontology/
%https://webprotege.stanford.edu/#projects/63bf402d-9c9c-4c61-b9c7-7af4b33ad6b0/edit/Classes?selection=Class(%3Chttp://webprotege.stanford.edu/RBGBmth90lfVxshJ8bfVbuO%3E)
%http://www.cs.ox.ac.uk/people/ian.horrocks/Seminars/download/Horrocks_Ian_pt1.pdf
%https://online.visual-paradigm.com/diagrams.jsp#diagramlist:proj=0

\section{Network Slicing: background and definitions}
\label{related}

Network slicing is being widely studied in Academia and Industry since it is a promising concept and technique for customizing and dedicating services according to client needs based on virtualization. Customizing services and isolating networks is actually not new. We can for instance refer to label-based protocols that encapsulates traffic into virtual links, such as \gls{MPLS} or IP tunnel-based paths performed by \glspl{VPN} among various technologies allowing differentiated quality of service management. 

The major innovation of network slicing comes together with NFV technologies. Both slicing and NFV enable customizing and deploying new network services on the fly while guarantee elasticity, scalability, and cost-efficiency. In other words, connectivity and specific treatments of traffic (both signaling and data) can be tailored to specific needs on demand. This claim enables operators to offer new services into an evolved framework of business models. 

A network slice has been defined in the literature in many different ways:  a bundle of services, a logical network, a type of virtual networking architecture, a chain of network functions created on top of a cloud infrastructure, etc. 

According to the GSM Association~\cite{SliceGSMA}, from a mobile operator point of view, a network slice is an independent end-to-end logical network on the top of  a shared physical infrastructure with a negotiated QoS. 

For Industry (see  \cite{SliceEricsson} for instance), network slicing creates separated \emph{use-case-specific logical networks} upon a shared physical infrastructure. This is enabled  by technology advances such as  \gls{NFV}, \gls{SDN} and \gls{MANO}. Network slices share the same physical network infrastructure but are effectively distinct and isolated up to some extent.

China Mobile, Huawei, Deutsche Telekom, and Volkswagen have released their shared vision for the 5G era~\cite{SliceHuawei}. In  their point of view, network slices are sets of dedicated logical Network-as-a-Service \gls{NaaS} meeting requirements coming from vertical industries. Through flexible and customized design of functions, isolation mechanisms, and O\&M tools, network slicing is capable of  providing logical dedicated networks upon a common infrastructure.

For China Telecom~\cite{SliceChinaTelecom}, a network slice is an end-to-end logical subnet requiring the coordination of a core network, an access, and a transport network. Network slices can be based on  isolated resources and/or shared resources. A network slice  is deployed in a service-oriented architecture enabling network features configurations (in terms of QoS, system capacity, data rate).

Other definitions can be found in the technical literature, see for instance those presented by European projects~\cite{5Gex,sliceNet}. It turns out that a network slice is commonly viewed as a logical network embedding network functions and IT resources in order to meet specific business requirements by a customer.

In this context, a \textit{network slice} is properly speaking not a service in itself; a slice  is invoked  to realize one or more services~\cite{SliceEricsson}. A network slice is then a logical network serving a defined business purpose with specific characteristics, and comprises all the required network resources. Such resources can be physical or virtual, and either dedicated to a particular slice, or shared between several slices~\cite{SliceEricsson}. The ETSI GR NGP in its Network Slicing Reference Framework~\cite{SliceETSI} defines a network slice as a description of a service aware logical network that is composed of different physical or virtual network elements, resources and functions. 

In the following section, to capture the complexity of network slicing, we use the concept of ontology. This subsequently allows us to come up with a model compatible with ONAP. The benefit of using an ontology approach is in that several models can be elaborated (possibly one model per operator) and are eventually composable in a global framework, which gives an enriched knowledge basis for formal reasoning. 

\section{Slice ontology}
\label{ontology}

\begin{figure}[hbtp]
    \centering
  \includegraphics[scale=\tallafigura, trim=350 60 0 0, clip] {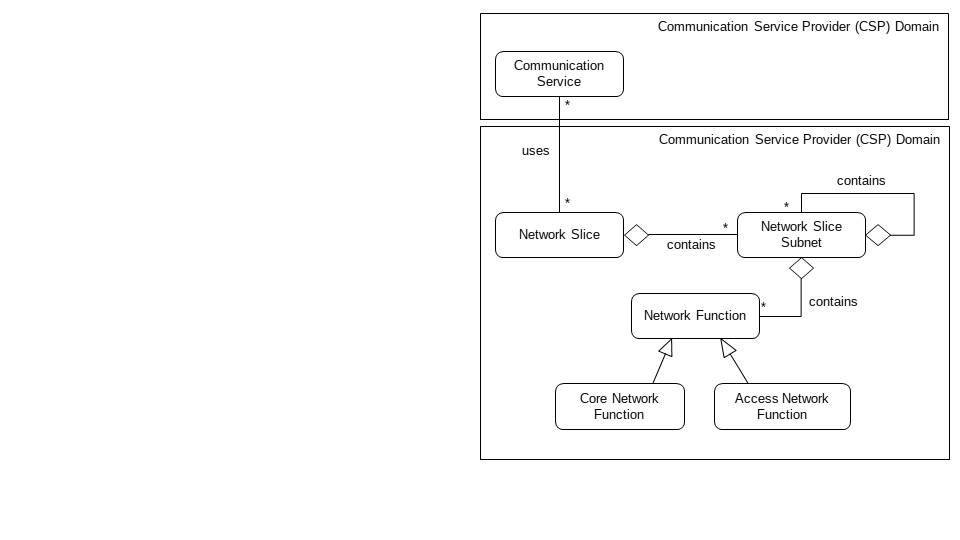}
    \caption{Network slicing ontology proposed by ONAP~\cite{SliceONAP}.}
    \label{onap}
\end{figure}

Various network slicing ontologies are available in the literature, see for instance that proposed in~\cite{celdran2018policy}. Another interesting ontology is given by the ONAP community in the framework of the 5G slicing use case~\cite{SliceONAP}. The model is based on 3GPP specifications and more concretely on the 3GPP TS.530~\cite{Slice3GPP530} (see Figure~\ref{onap} for an illustration).

This ontology notably complies with the 5G architecture defined in 3GPP TS 23.501 \cite{3GPP501}. It considers that communication service instances are provided by various \glspl{NSI}. In addition, it introduces the concept of network slice subnet, which is used by both  3GPP and ETSI. A complete slice is then the interconnection of slice subnets referred to as \glspl{NSSI}. A slice subnet is in turn composed of network functions. According to  ETSI a subnet represents single or multiple networks under the control of an agent~\cite{SliceETSI}. An agent makes reference to a logical entity that has complete control of its network infrastructure. 

\begin{figure}[hbtp]
    \centering
  \includegraphics[scale=\tallafigura, trim=330 130 0 0, clip] {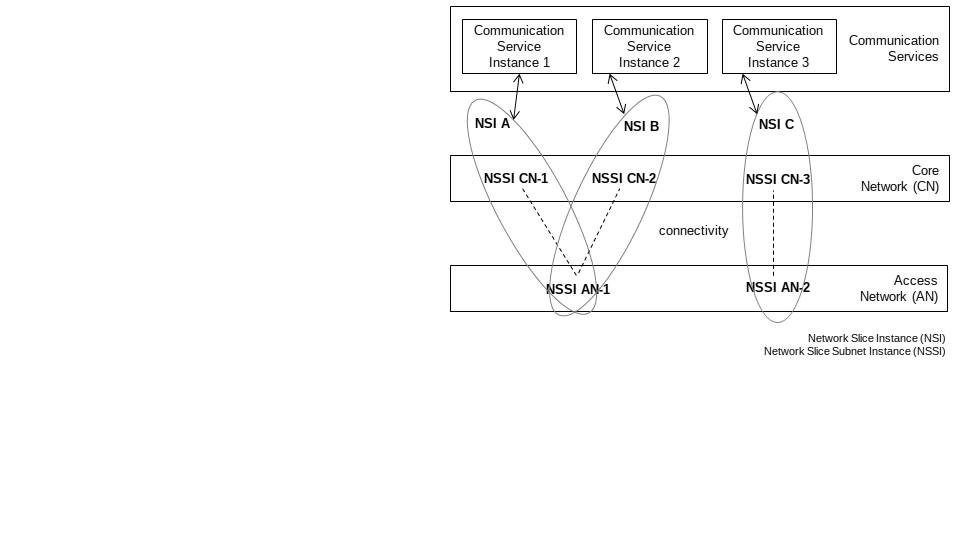}
    \caption{An example of Network Slicing~\cite{Slice3GPP530}.}
    \label{onap1}
\end{figure}

Figure~\ref{onap1}  shows that different slices ($A$, $B$ and $C$) contain network functions belonging to the \gls{NSSI} Access and Core Network.  Communication Services are then provided through different \glspl{NSI}~\cite{Slice3GPP530}. The above ontology is purely descriptive and does not consider the hosting infrastructure, which shall enable  the deployment  of VNFs belonging to slices. In fact, the performance of network slices directly depends of VNF placement that compose them. For instance, slices requiring low latency performance need to be placed as close as possible to end users, network functions executing computation intensive tasks require high performance processors, etc. Thus, it is possible to develop another ontology taking into account all the processes involved in the creation of a slice. The goal is to develop a method of reasoning when dealing with multi-domain cases, i.e. when several actors (or operators) are involved in the creation of a slice. 

For this purpose, we can consider the creation of a slice as a deployment exercise~\cite{ONISTT1}. 

This approach aims at guaranteeing that the capabilities provided by systems can fulfill the needed capabilities. 

The \gls{ONISTT} approach is presented in~\cite{celdran2018policy}. The authors use \gls{OWL} to express different types of ontology. 

\begin{figure}[hbtp]
    \centering
  \includegraphics[scale=\tallafigura, trim=300 50 0 0, clip] {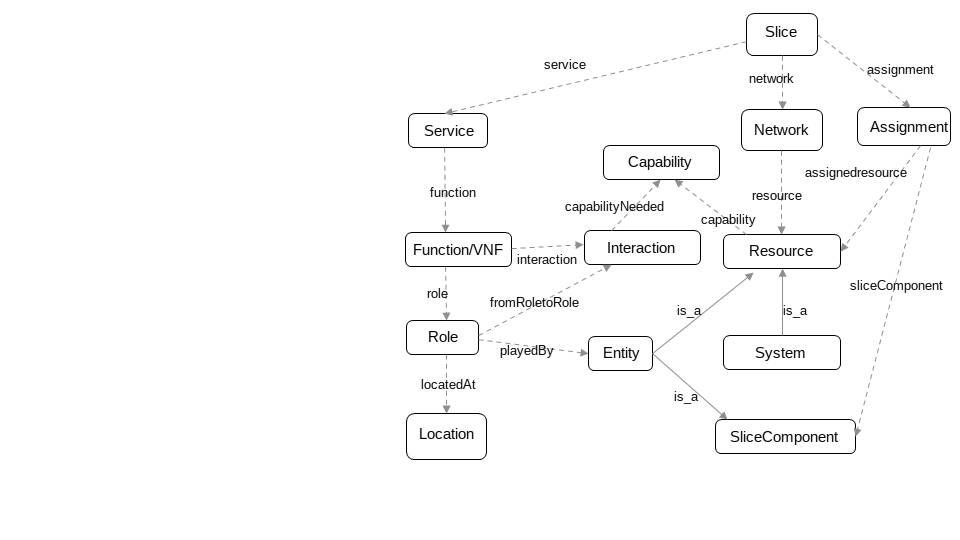}
    \caption{ONISTT-based network slicing ontology.}
    \label{bundlecomponents}
\end{figure}

The strength of \gls{ONISTT} is that it consider: \textit{(i)} an analyzer that applies general logical reasoning and domain-specific rules to determine whether a candidate confederation can satisfy the requirements of a proposed event; \textit{(ii)} the knowledge is captured from instance data into a \gls{KB}. It turns out that \gls{ONISTT} ontology concepts can be adapted to the context of  network slicing. A top-level ONISTT-based ontology is shown in Figure~\ref{bundlecomponents}.

To adapt the ONISTT approach to slicing, we use the generic term \network to design the classical network.
That is a collection of network elements such as switches, routers, optical cross-connects, etc. 

interconnected by transmission links and possibly controlled by SDN controllers) augmented by IT resources (disk, CPU, RAM) hosted in data centers. The slice ontology has three complementary parts that compose a \deployment (i.e., when instantiating a network slice):
\begin{itemize}
    \item A \bundle has \service objectives (e.g., an interconnection between origin-destination pairs such a classical VPN, a virtual EPC, etc.), from which an assemblage of capabilities is needed to offer all individual services composing the slice bundle.
    \item A \network is a set of \resources offering individual capabilities, which can be combined to support the capabilities needed by a \service.
    \item A set of \assignments match the capabilities offered by individual network resources with those required to support the service bundle.
\end{itemize}

The resulting ontology is illustrated in Figure~\ref{bundlecomponents} and covers all the aspects of a slice deployment. In the following section, by keeping in mind all the aspects of slice deployment, we focus on an ontology which is compatible with ONAP.

%While the drawback of the ONAP approach is that it does not consider the infrastructure hosting the VNFs, the ONISTT-based solutions seems to be more complete but more complex as well. 

\section{Unified Network Slicing Model}
\label{onap_ontology}

\begin{figure*}[hbtp]
    \centering
  \includegraphics[scale=\tallafigurabig, trim=0 0 0 0, clip] {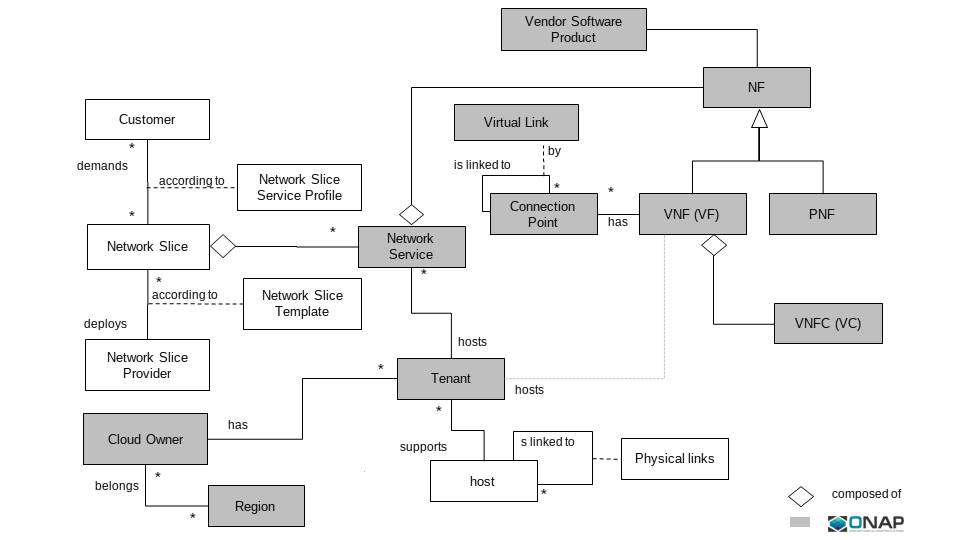}
    \caption{ONAP-compatible network slicing ontology.}
    \label{SliceOnt}
\end{figure*}

In this section, we continue along the line of defining a network slicing ontology by keeping in mind that it has to be compatible with the automation platform that shall orchestrate the network slices (namely, ONAP). The proposed model considers the required entities, actors and properties for automating the deployment of network slices. It is shown in Figure~\ref{SliceOnt}. Our proposal  not only considers the slicing principles introduced by both 3GPP~\cite{Slice3GPP530} and ETSI~\cite{SliceETSI} but moreover the deployment requirements when considering the automation of the whole lifecycle of network slices, and more specifically when using \gls{ONAP}.

We begin considering that a slice is by definition composed of a bundle of services offered by a single or several virtualized network functions running on shared or dedicated infrastructures. In general, a slice, beyond its polymorphic nature, is composed of a number of services to meet the global slice objectives/requirements. These requirements can be specified in a slice \gls{SLA} detailing the objectives for the various services composing the slice (referred to as Network Slice Service Profile); each service can even have its own \gls{SLA} and the \glspl{SLA} of the various services composing a slice can be combined to achieve the \gls{SLA} of the slice.

The proposed model involves the following entities:
\begin{itemize}
    \item \textit{Customer}: Defines the customer properties (e.g., name, description, etc.). A customer is the entity that requires the deployment of a network slice and shall use it. A customer can be an enterprise, a single user, or a group of users belonging to a category (e.g., e-Health,). 
    
    \item \textit{Network Slice Provider (NSP)}: Deploys a network slice on the basis of network resources or services from multi-domain and/or multi-technology networks~\cite{SliceETSI}.
    
       \item \textit{Vendor Software Product}: This is the owner of an ONAP-resource. The defined ONAP resources are: \gls{VF}, \gls{VFC}, \gls{CR}, \gls{PNF}, \gls{CP} \gls{VL}.

    %\item \textbf{Provider}: Entity that supplies  a service, a software product implementing a VNF, a PNF,  a network slice and/or a cloud infrastructure.
    
    \item \textit{Network Service}: Bundle of network functions that accomplishes a customer need. The granularity of a network service is given by the designer. For instance an end-to-end 5G network can be seen as a single service; however, both RAN and Core networks can be also viewed as individual services, namely RANaaS, 5GCNaaS. The final goal is to achieve  loosely coupled services in order to improve modularity and individual deployment. Smaller services enable flexible, efficient and agile developments and deployments. The management of services become also more resilient. 
    
    \item \textit{Network Slice}: Logical entity that is composed by a single or multiple network services (also referred to as subnets). When using the concept of \textit{slice subnet}, a \gls{NSS} carries out a single service. In the following, we avoid using the subnet concept because it introduces unnecessary complexity and ambiguity. The granularity of a given service and its components shall generally be defined by the software product vendor who shall accomplish given guidelines (e.g., by network operators).
    
    %A slice is composed by network services, which in turn are composed by subservices (referred to as services). 
    
    \item \textit{Network Function}: Entity that accomplishes a task for a given network service. A network function can be a \gls{PNF} or a \gls{VNF}. A network function can in turn be composed of network sub-functions. The design of network functions and their components (referred to as VNFC) shall be determined by the software product provider.

     \item \textit{Network Slice Service Profile}: Contains the description of a customized service. From a business point of view, service profiles constitute the available \textit{slice offers} that are or can be commercialized. A \gls{NSSP} can then be offered to a potential slicing consumer as a basis on which clients can customize their slices. Certainly, offering standard \glspl{NSSP} shall be enough to satisfy the majority of customers. Examples of network slice requirements that can be part of the {NSSP} are: area traffic capacity, charging, coverage area, degree of isolation, end-to-end latency, mobility, overall user density, priority, service availability, service reliability, UE speed, among others~\cite{Slice3GPP530}.

    \item \textit{Network Slice Template}: Technical description of a network slice. It contains attributes values for the creation and management of network slice. For a given network slice, the Slice Template specifies the charging capacity, guaranteed latency, guaranteed reliability, guaranteed data rate as well as the resource requirements (CPU, RAM, storage, network) for each network function composing the network service. 
    
    A slice Template is then directly related to HEAT files describing the network services that compose the underlying slice.
    
        \item \textit{\gls{SLA}}: Contract  established between a customer and the Network Slice provider according to the requirements defined in the NSSP and that shall be deployed under the principles of the Slice Template.

   % \item \textit{NF provider}: Designs, implements, and offers physical and/or virtual network functions.
    
    \item \textit{Physical link}: Entity that represents physical connections between hosts, which can be located in distant geographic zones. According to ETSI~\cite{SliceETSI} network links are a type of resource.

    \item \textit{Host}: Physical compute engine, which can support a single or multiple tenants. 
    
    \item \textit{Tenant}: Isolated set of resources (compute, network, storage) belonging to a customer or to a network slice provider.

    \item \textit{Virtual link}: Enable the connection of VNFs via  connection points. 
    
    \item \textit{Connection Points}: Represents the network ports of VNFs. Connections points enable communicating two or more virtual network functions.

\end{itemize}

    Virtual Links and Connection Points are rarely used. In fact, network ports, virtual networks and subnetworks are generally defined as part of network services or network functions. In ONAP, a network service is usually defined by a single HEAT template or composed of a forwarding graph of available network functions. 
    
  The various elements of the proposed slice ontology are illustrated in Figure~\ref{SliceOnt}. The grey boxes indicate those covered by ONAP. As we can observe many elements are missing in ONAP in order to have a holistic view of network slicing within the automation platform. We illustrate the above ontology in by a mobile core network.

\section{Network Slicing Automation: Illustration for a customized mobile core network}
\label{example}

\subsection{Use case description}

We consider the case of a company willing to deploy its own mobile core network on the top of a commercial mobile network, i.e., by using the radio access network of a network operator (say Grey Operator). The private core network (referred to as Slice A) can then implement customized policies in terms of security, reliability and quality of service (latency, data rate, etc).  

\begin{figure}[hbtp]
  \centering
   \includegraphics[scale=0.31, trim=640 10 0 0, clip] {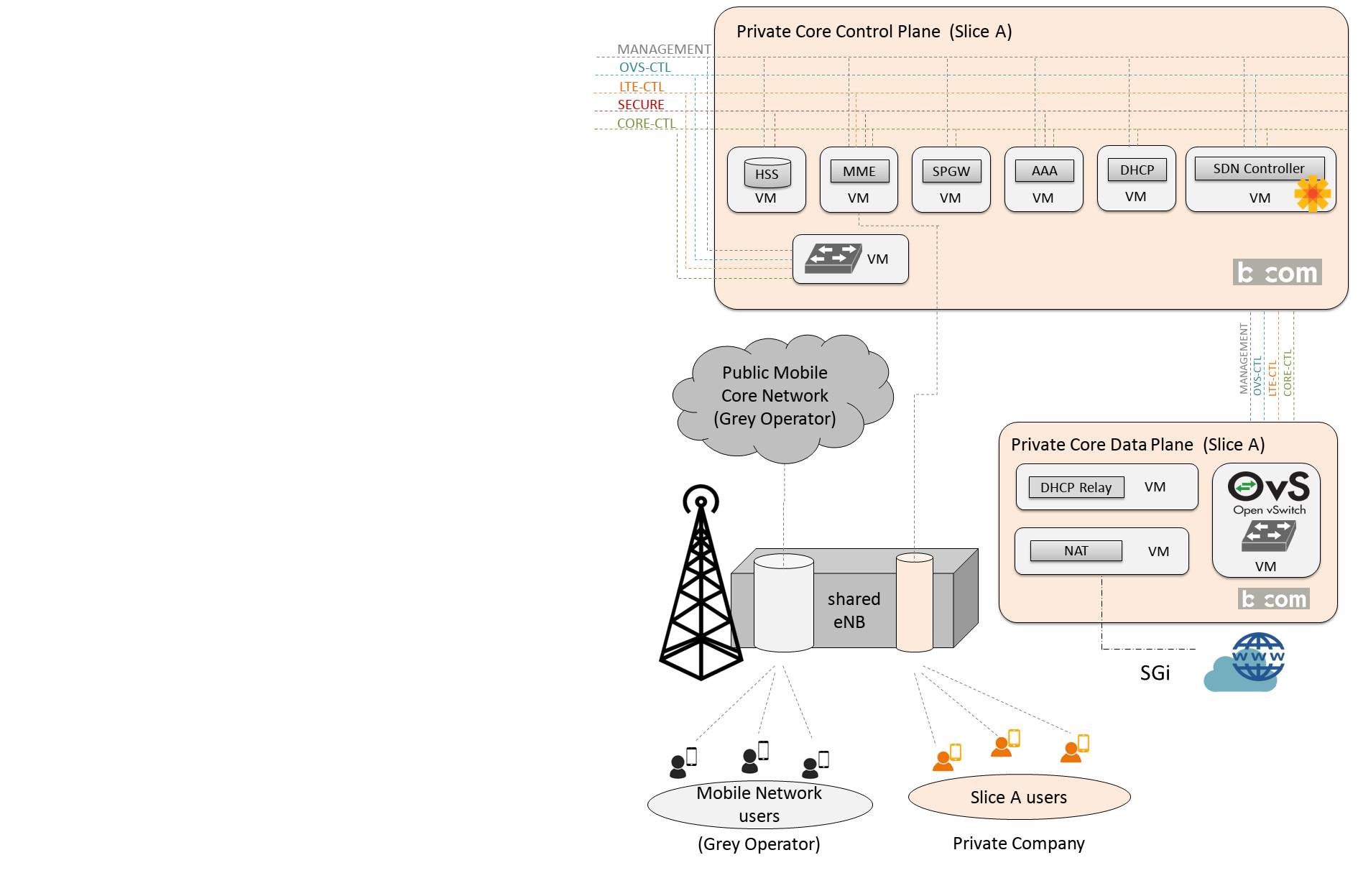}
       \caption{Network slicing: A private and customized mobile core network.}
  \label{fig:wef}
\end{figure}

The shared eNodeB is in charge of rerouting the private and public traffic to the adequate core network (namely, to the corresponding MME in 4G networks); see Figure~\ref{fig:wef} for an illustration. A routing solution has been proposed by the \textit{eDecor} project~\cite{Decor}; we hence consider this issue as being solved. In the following, we focus on the network slice deployment, i.e., the onboarding and  instantiation of a private mobile core network by using a virtualized infrastructure managed by ONAP. 

The proposed network slice specifically implements a customized mobile core network based on various open-source solutions, notably \gls{OAI} code. The mobile core network considered in this paper is a fully virtualized 4G core network. 

We specifically use the virtualized 4G core network implemented by b$<>$com’s solution~\cite{bcom}. It notably realizes  a complete separation of the user and control planes as recommended by 3GPP for 5G networks, referred to as \gls{CUPS}. The various components of the core network are connected via VxLANs (namely, MANAGEMENT, OVS-CTL, LTE-CTL,SECURE, CORE-CTL). 

When a UE attaches to the private core network, the \gls{AAA}  procedure is triggered by the \gls{MME}. User profiles are validated by the \gls{HSS} data base which stores various parameters such as apn, imei, sim-card information OP, key, \gls{MNC}, \gls{MCC}, among others.  When access is granted to the UE, the DHCP component provides it the IP-address, which is taken out of the address-pool established in the AAA component. The end-to-end connection is assured after the creation of the GTP-U and GTP-C tunnels. The NAT component provides address translation and is deployed between the SGi interface and the Internet network.

\subsection{Testbed settings}

The onboarding and instantiation of the private mobile core network (Slice A) has been carried out with ONAP Casablanca~\footnote{\gls{ONAP} is the result of the merge of two open source projects: Open \gls{ECOMP} from AT\&T  and Open-O from Linux foundation \gls{MANO} project. The ONAP project was formed in March, 2017 in response to a rising need for a common platform for telecommunication, cable, and cloud operators—and their solution providers—to deliver differentiated network services on demand, profitably and competitively, while leveraging existing investments~\cite{onapIntro}.}. We concretely create three private tenants on an Openstack-based platform. 

\begin{figure}[hbtp]
  \centering
   \includegraphics[scale=\tallafigura, trim=300 140 0 0, clip] {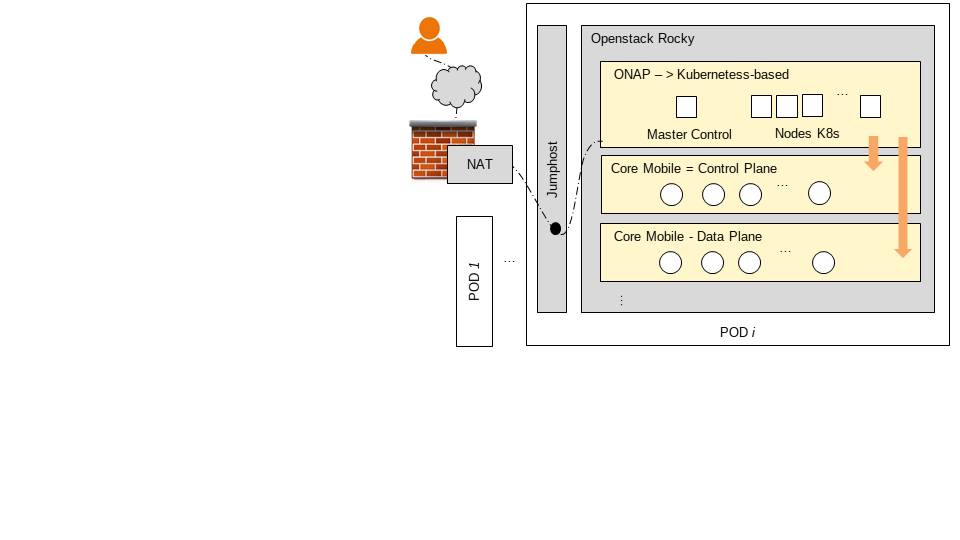}
    \caption{ Network Slicing Automation: Testbed architecture.}
  \label{fig:testbed}
\end{figure}

The first tenant contains the ONAP subsystems and the two others are respectively reserved for the deployment of the control plane and user plane of the private core network. The testbed architecture is shown in Figure~\ref{fig:testbed}. The ONAP platform provides independent subsystems to design, create and manage the life-cycle of network services. Network slices can be seen as network services. 

%The main components of both design- and run-time frameworks are illustrated in Figure~\ref{fig:onap_components}.  In addition, integration and validation utilities are available for automating these procedures, i.e.,the \gls{VVP} and the `Onap-tests' modules. This latter enables automatically onboarding, distributing and instantiating VFs on ONAP.

%\begin{figure*}[hbtp]
%  \centering
%   \includegraphics[scale=0.5, trim=0 0 160 0, clip] {Diapositive1.JPG}
%  
%    \caption{ONAP Subsystems.}
%  \label{fig:onap_components}
%\end{figure*}

\subsection{Network Slice design}

Generally speaking, ONAP considers a Network Service as a chain of \gls{VNF} (also referred to as \gls{VF}) which in turn can be formed by \glspl{VNFC}. We define the Network Slice A (private mobile core network) as a chain of two independent services corresponding to the control and data plane functions. Each service is formed by a single VF, i.e., Core-CP-VF (control plane), Core-DP-VF (data plane). Each VF is in turn formed by various components.  VFs are described by means of Heat Templates (one for each) and their components correspond to Openstack resources (OS::Nova::Server, OS::Neutron::Net,OS::Neutron::Subnet,OS::Neutron::Port). 

\begin{figure}[hbtp]
  \centering
   \includegraphics[scale=\tallafigura, trim=290 0 0 0, clip] {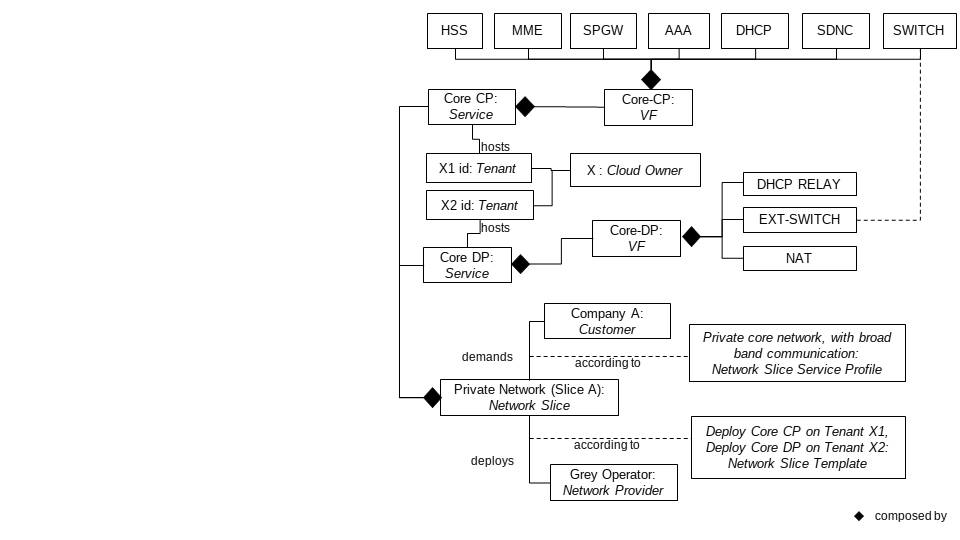}
   
    \caption{ONAP-based mobile core network model.}
  \label{fig:service_wef}
\end{figure}

The resulting service model to be deployed on ONAP is shown in Figure~\ref{fig:service_wef}. The \textit{Grey Operator} is then the Network Provider that deploys the Slice A demanded by a given company. Slice requirements are defined by the customer and included in the Network Slice Service Profile. The customer requirements are captured by the Network Slice Template which defines the technical aspects that enable deploying the demanded slice (e.g., network services, cloud owner, tenants, etc).

\subsection{Slice deployment workflow}
In ONAP, the process for readying a service for distribution involves various roles based on a complex workflow. Before instantiating a network slice (an ONAP service) various stages need to be performed. They include onboarding, approving and deploying procedures which belong to the design-time environment of ONAP. The automation platform also supports run-time functions such as policy-driven automation and closed-loop management; these functions are out of the scope of the present paper. 

For dealing with the various stages of the service onboarding and deployment, ONAP defines various roles: superuser, tester, designer, governor, and operator. 

\begin{figure}[hbtp]
  \centering
   \includegraphics[scale=\tallafigura, trim=350 60 0 0, clip] {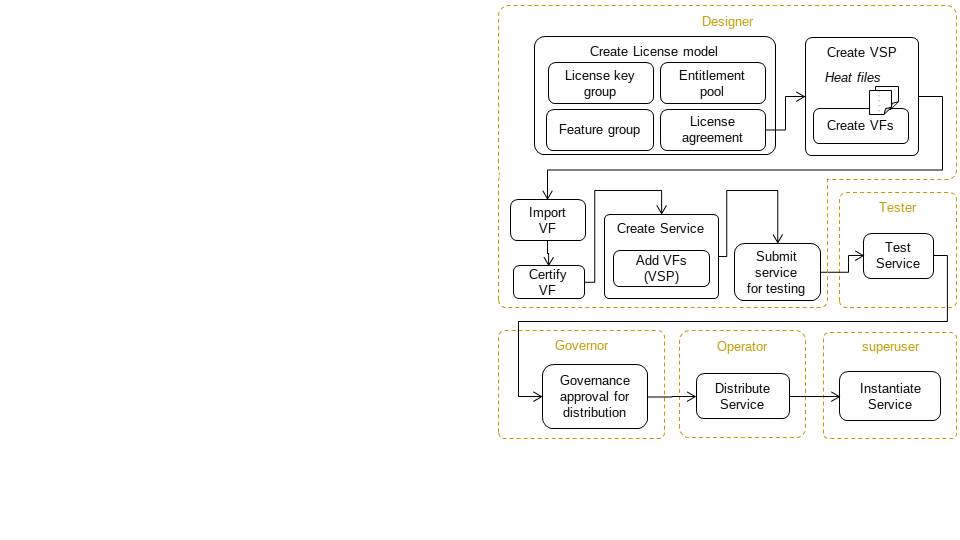}
   
    \caption{ONAP-based service deployment workflow.}
  \label{fig:lifecycle}
\end{figure}

The workflow of a service deployment with ONAP is shown in Figure~\ref{fig:lifecycle}. The whole workflow is carried out by each service composing the network slice. Since the proposed model for the private mobile core network slice involves two services (Core-CP and Core-DP), the workflow has been twice performed.  

The slice deployment begins with the validation of \glspl{VF}. We assume that the images of the various \glspl{VFC} exist. In practice, these images are created by the VNF provider. They can be in the form of \texttt{.qcow2} in order to upload them onto an OpenStack platform. This latter task requires the preparation of Heat templates, which need to be ONAP-compatible (See Section~\ref{heattem} for details).

Thus, we build the ONAP-compatible HEAT templates for both Core Control-Plane and Core Data-Plane functions.  Then, we create and certify them as ONAP \glspl{VF}. For realizing the Core-CP-VF and Core-DP-VF for instantiation, we create, test and distribute the corresponding services. Finally, we instantiate the Core-CP and Core-DP services through the \gls{VID} app of ONAP.

\subsection{Heat Template guidelines}
\label{heattem}

 ONAP defines stringent naming conventions and restricts the use of various Openstack resources such as Resource~OS::Neutron::FloatingIP, Resource~OS::Neutron::FloatingIPAssociation, among others. The ONAP requirements concerning the definition of HEAT files are detailed in~\cite{onap_requirements}. In addition, ONAP requires mandatory metadata when defining nova-server resources. Metadata must include the following parameters: \textit{vnf\_name}, \textit{vnf\_id}, \textit{vf\_module\_id}. 
 
 On the other hand, the ONAP environment file must not be greater than $2000$ characters (including quotes for each of them). In fact ONAP concatenates all environment variables to be inserted as a single database field (namely, \textit{Resource\_input}) into the \textit{vnf\_resource\_customization} table during the service onboarding. This size will be upgraded to $20000$ in the Dublin version of ONAP, while for Casablanca a patch is required.

%

%for dealing with the various subsystems according to the service lifecycle (design, testing, approval, distribution, instantiation, scaling).

%and involve~\cite{design_onap}:

%\subsubsection*{Onboarding}
%The goal of the onboarding phase is to add models and other artifacts required to create, configure, instantiate and manage \glspl{VNF}. This stage is carried out by the designer by means of the \gls{SDC}. The onboarding includes:
%\begin{itemize}
%    \item License model creation
%    \item \acrfull{VSP} creation
%    \item \acrfull{VF} creation: VFs are the building blocks of a service
%    \item Importing VFs
%    \item Certifying VFs
%\end{itemize}

%\subsubsection*{Service design}
%\begin{itemize}
%    \item Service Creation
%    \item Service composition from available VFs
%    \item Submit service for testing
%\end{itemize}

%\subsubsection*{Testing and approvals}
% This step is carried out by the \textit{tester} who validates the service and submit it for approval. Once the \textit{governor} approves the service, it is thus ready for distribution. The \textit{operator} role distribute the service, leaving it available for instantiation. 

%\subsection*{Instantiation}
%The instantiation stage deploys the service. This action is performed by the \textit{superuser}. 

\subsection{Gap analysis}

In ONAP, a network slice might be defined as one or more network services. However, an upper abstraction for managing these various network services that compose a network slice has not been defined so far. When a single network slice is defined by more than one network service (e.g., the proposed private core network which is composed of two services, Core-CP and Core-DP) the lifecycle management of each of them need to be  performed independently. 

The main drawback of defining a network slice as a network service is that the whole service shall be instantiated on a single hosting infrastructure (namely, an Openstack tenant). By definition in ONAP, a network service represents the finest granularity of placement. In other words, all VFs belonging to a network service shall be instantiated on the same tenant. 

When considering the end-to-end performance (e.g., in terms of latency) of a given network service, some of their components (VFs) may require to be placed at different geographic locations.  Ideally, VNFs should be located, where they are the most efficient in terms of performance and cost. The problem of VNF placement has been widely studied in the literature ~\cite{luizelli2015piecing,slim2017towards} in order to guarantee service requirements while optimizing cloud and network resources (i.e., bandwidth computing, memory, storage). 

Even though the ONAP community has defined the \gls{OOF} module for dealing with placement, these functionality has not been fully exploited.  

%Beyond placement, the performance of a network service is also determined by the computing capacity and system architecture of the hosting infrastructure. Virtualized network functions requiring real-time behavior take advantage of container-based architectures. Unlike VM-based systems, which require a complete OS for running processes, containers require only the associated libraries. This  makes containers  lighter than VMs, and also faster and more suitable when deploying low latency services. However, ONAP actually does not enable container-based deployments. ONAP-Casablanca only supports Openstack infrastructures, where VNFs are instantiated on the top of VMs. ONAP-Dublin promises including  container-based infrastructures (namely, Kubernetes), however this functionality shall be probably mature in ONAP-El-Alto. %Heterogeneous environments shall allow network and cloud operators to optimize the service lifecycle automation as well as their physical and virtual infrastructures. 

%ONAP includes multi-site and multi-VIM orchestration capabilities for \glspl{VNF} and also enables full lifecycle management for \glspl{PNF}. ONAP addresses the service lifecycle automation in order to avoid the cost of manual changes currently required to implement new service offerings. In addition ONAP seeks to exploit \gls{NFV} and \gls{SDN} advantages. The final goal is to achieve both cost reduction and network performance increase. 

\section{Conclusions}
\label{conclusion}

%Network slicing is the core of future mobile networks (5G and beyond). It enables addressing new services and new requirements according to the needs by vertical markets, such as real-time capabilities, reliability and guaranteed Service Level Agreements (SLAs). In this paper, we propose an ONAP-compatible network slicing ontology which enables defining network slices to be deployed with an automation platform as ONAP. The proposed model puts together both ONAP software entities and 3GPP network architectures. As driving usecase of this study, we concretely define a network slice that implements a private and customized mobile core network that aim to be dedicated to a given company.

We have proposed a network slicing ontology  compatible with ONAP. The proposed ontology takes into account major aspects of network slice deployment that we have illustrated by considering a network slice based on a private core network. This has enabled us to identify some shortcomings in the development of ONAP. 

It turns out that in order to reach end-to-end performance requirements as well as greater flexibility and scalability, the components belonging to a network slice may require to be instantiated on different geographic locations (namely, various Openstack tenants or platforms) but seen by ONAP as a unique object. This in fact requires to introduce an intermediate abstraction layer for managing the various network services that compose a network slice. This abstraction layer should also enable selecting the hosting infrastructure of each network service component according to performance-based placement policies. 

% to hide the heterogeneity of the underlying virtualized infrastructures.

%We expose the various stages involved in the slice deployment, i.e. design, onboarding, instantiation and distribution of both network services. Users connect to the private core network by means of a shared access network (namely, a commercial eNB) which redirects the traffic to the corresponding core network. We end this paper with a brief discussion on the ONAP limitations when deploying network slices. We notably consider that the muti-site and multi-VIM orchestration capabilities need to be improved in future ONAP releases in order to achieve a finest granularity in the placement of network service components as well as the support of container-based infrastructures. 

% Network Slicing behavior of 5G networks and their architecture as well as the automation of the slicing deployment by means of open-source platforms as ONAP. Finally, the audience shall be able to identify advantages and limitations of available open-source projects in the framework of 5G network slicing, NFV-SDN, management and orchestration. 

%The session shall expose the whole lifecycle of an open source-based 5G network slice, namely a virtualized mobile core and access  networks (with private antennas) that aim to be dedicated to a given company.

%Both cloud and network critical requirements shall be addressed on the basis of an unified network slicing model

%\appendix 

  \setglossarystyle{modsuper}
{\footnotesize
\printglossary[type=\acronymtype, nonumberlist]
}

\bibliographystyle{IEEEtran}
\bibliography{biblo}
\noindent

\end{document}